\documentclass[12pt]{iopart}
\usepackage{iopams}
\usepackage{graphicx} 
\usepackage{amsmath}
\usepackage{amssymb}
\usepackage{caption}
\usepackage[left, pagewise]{lineno}
\usepackage[dvipsnames]{xcolor}

\newcommand{\Trr}{\mathrm{Tr}}

\newcommand{\MM}{\mathsf{M}}

\newcommand{\trm}{\Trr(\MM)} 
\newcommand{\vv}{\mathbf{v}}
\newcommand{\xx}{\mathbf{x}}

\newcommand{\pcanorm}{\tilde{I}_\text{PCA}}
\newcommand{\pcbnorm}{\tilde{I}_\text{PCB}}

\newcommand{\ccnorm}{\tilde{I}_\text{CC}}
\newcommand{\fixed}{\text{(fixed point)}} 
\newcommand{\nfixed}{n_\fixed}
\begin{document}
\title[W7-X Edge Topology]{Characterisation of X- and O-points in Wendelstein 7-X with respect to coil currents}
\author{Robert Davies${}^1$, Christopher B. Smiet${}^2$, Charlotte Batzdorf${}^3$, J. Geiger${}^1$, J. Loizu${}^2$, S. A. Henneberg${}^4$}
\address{${}^1$Max Planck Institute for Plasma Physics, Wendelsteinstraße 1, 17491 Greifswald, Germany \\
${}^2$Ecole Polytechnique Fédérale de Lausanne (EPFL), Swiss Plasma Center (SPC), CH-1015 Lausanne, Switzerland  \\
${}^3$RWTH Aachen University, Aachen, Germany \\
${}^4$Massachusetts Institute of Technology, Cambridge, Massachusetts 02139, USA
}
\date{\today}
    
\begin{abstract}
This work analyses vacuum magnetic field topology in Wendelstein 7-X (W7-X) with respect to changes in the current in the superconducting coils. We develop a fast automated scheme to locate fixed points (such as X- and O-points) and calculate the trace of the Jacobian of the field line map for them ($\trm$), which represents several important properties of the fixed point. We perform two sets of coil current scans: (1) scans where each coil current is varied individually, using the ``standard", ``high iota" and ``low iota" configurations as starting points; (2) a scan of over $2\times10^5$ magnetic configurations in which the coil currents are randomly sampled. In both cases we constrain the coil currents to the normal range of W7-X. We verify the principal roles of the non-planar, planar and control coils: the non-planar coils establish island chains with a certain phase; the planar coils modify the location of the island chain by both controlling the $\iota$ profile and shifting the configuration ``inward" and ``outward"; the control coil affects the island size and phase. We also find that $\vert(\trm-2)\vert$ (a quantity closely related to the magnitude of the Greene's residue) tends to increase with the minor radius of the fixed points, and that $\trm$ for X- and O-points can be very differently affected by the control coil current. Finally, we show that $\vert(\trm-2)\vert$ serves as a proxy for island size for internal island chains, which may help  identification of suitable experimental candidates.
\end{abstract}
\maketitle

\section{Introduction}
\begin{figure}
    \centering
    \includegraphics[width=0.5\linewidth]{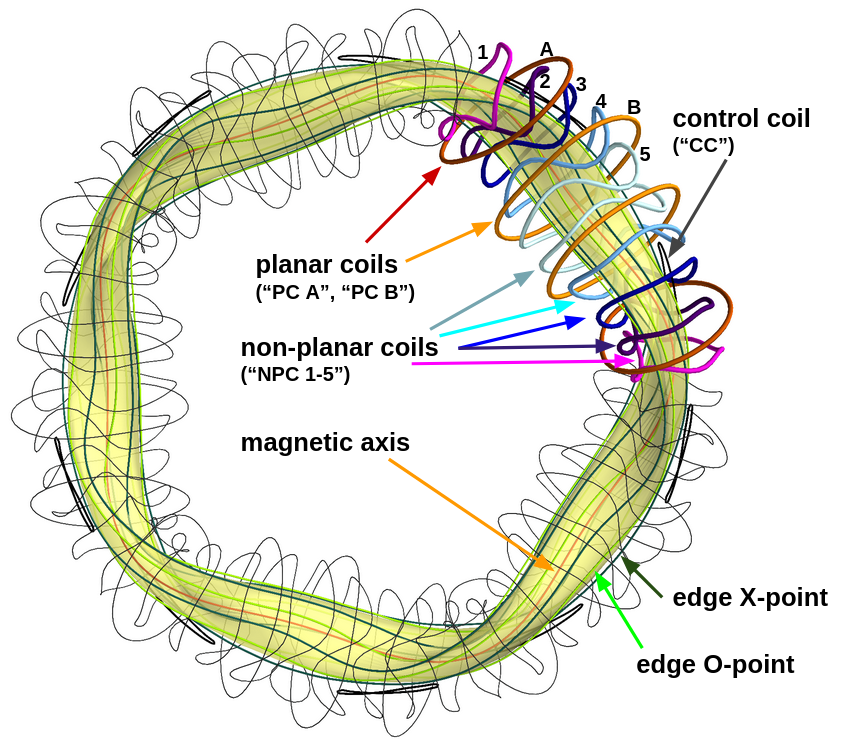}  
    \caption{Top-down view of W7-X for the ``standard configuration". The plasma boundary is shown as a yellow surface and coils as black lines, with the coils of one field period shown bold and coloured (each unique coil geometry having its own colour). The magnetic axis is shown as a red line and the X- and O-points of the edge island chain are shown as dark and light green lines.}
    \label{fig:coils_plot}
\end{figure}
\begin{figure}
    \centering
    \includegraphics[width=0.48\linewidth]{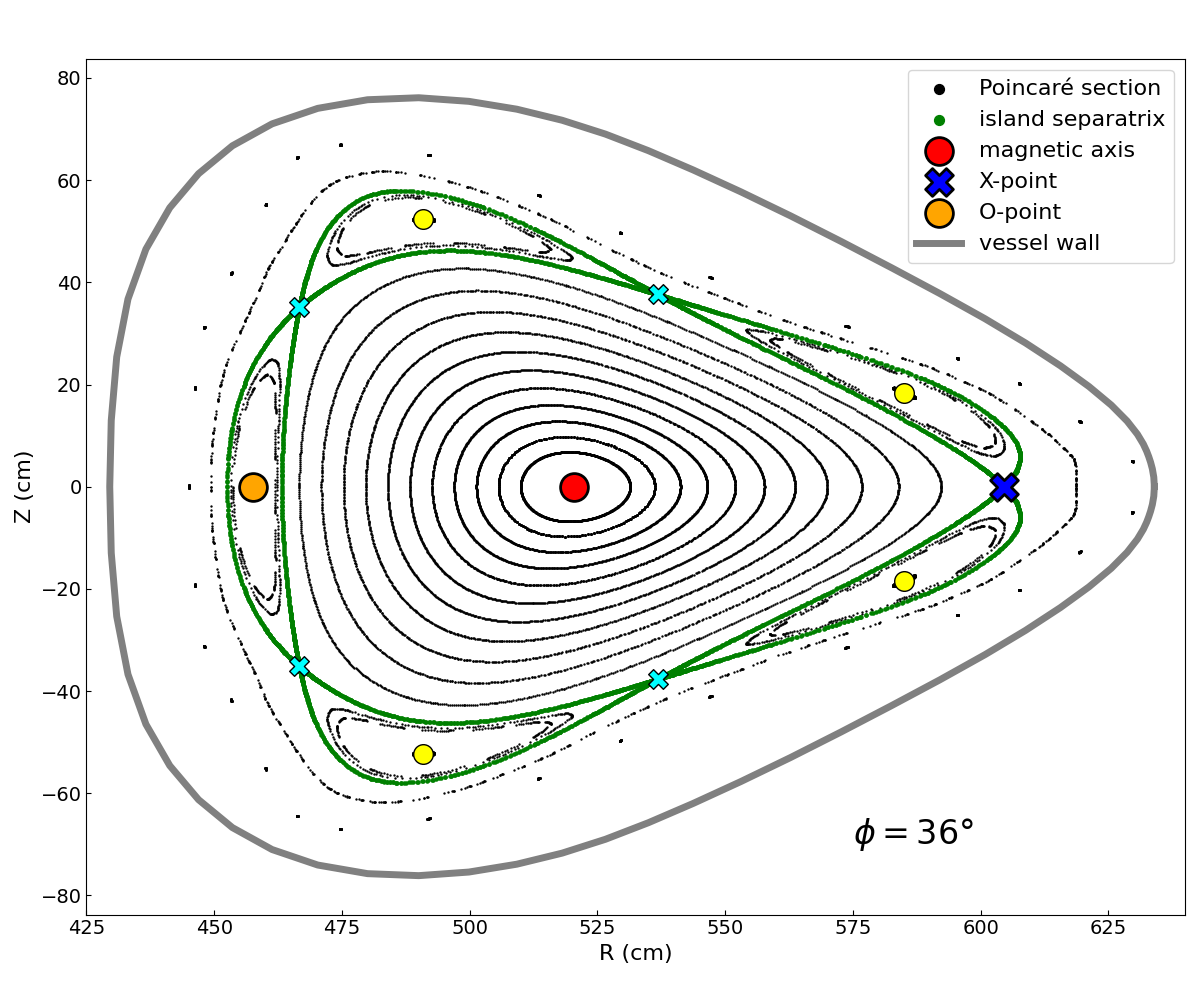}    
    \includegraphics[width=0.48\linewidth]{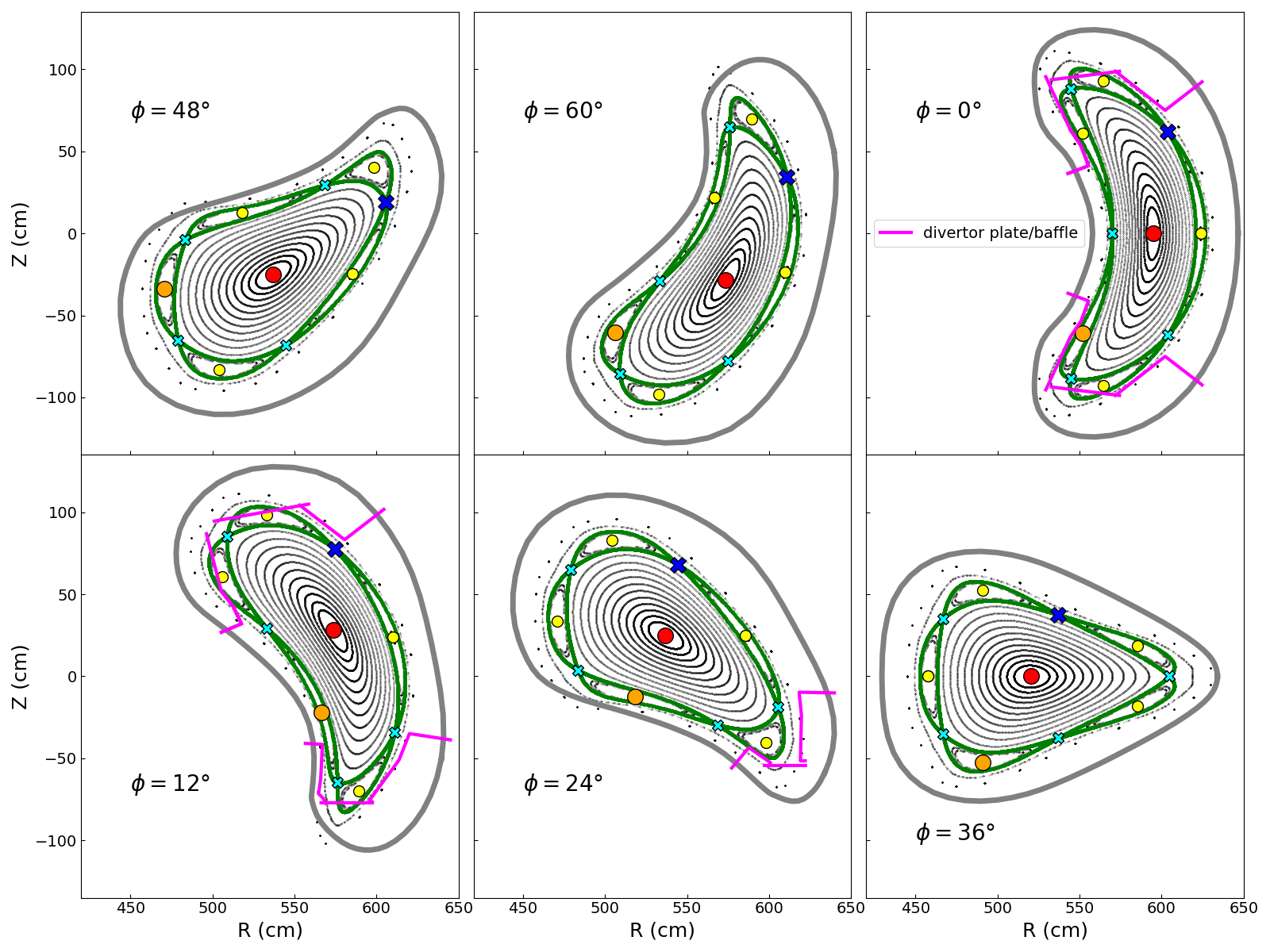}    
    \caption{A Poincar\'e section for the ``standard" configuration at several toroidal locations, also showing the magnetic axis and edge $\iota=1$ X- and O-points. The X-point and O-point in the midplane ($Z=0$) at $\phi=36^\circ$ are shown by larger markers (blue and orange) to illustrate their motion.}
    \label{fig:standard_explanatory}
\end{figure}
\begin{figure}
    \centering
    \includegraphics[width=0.6\linewidth]{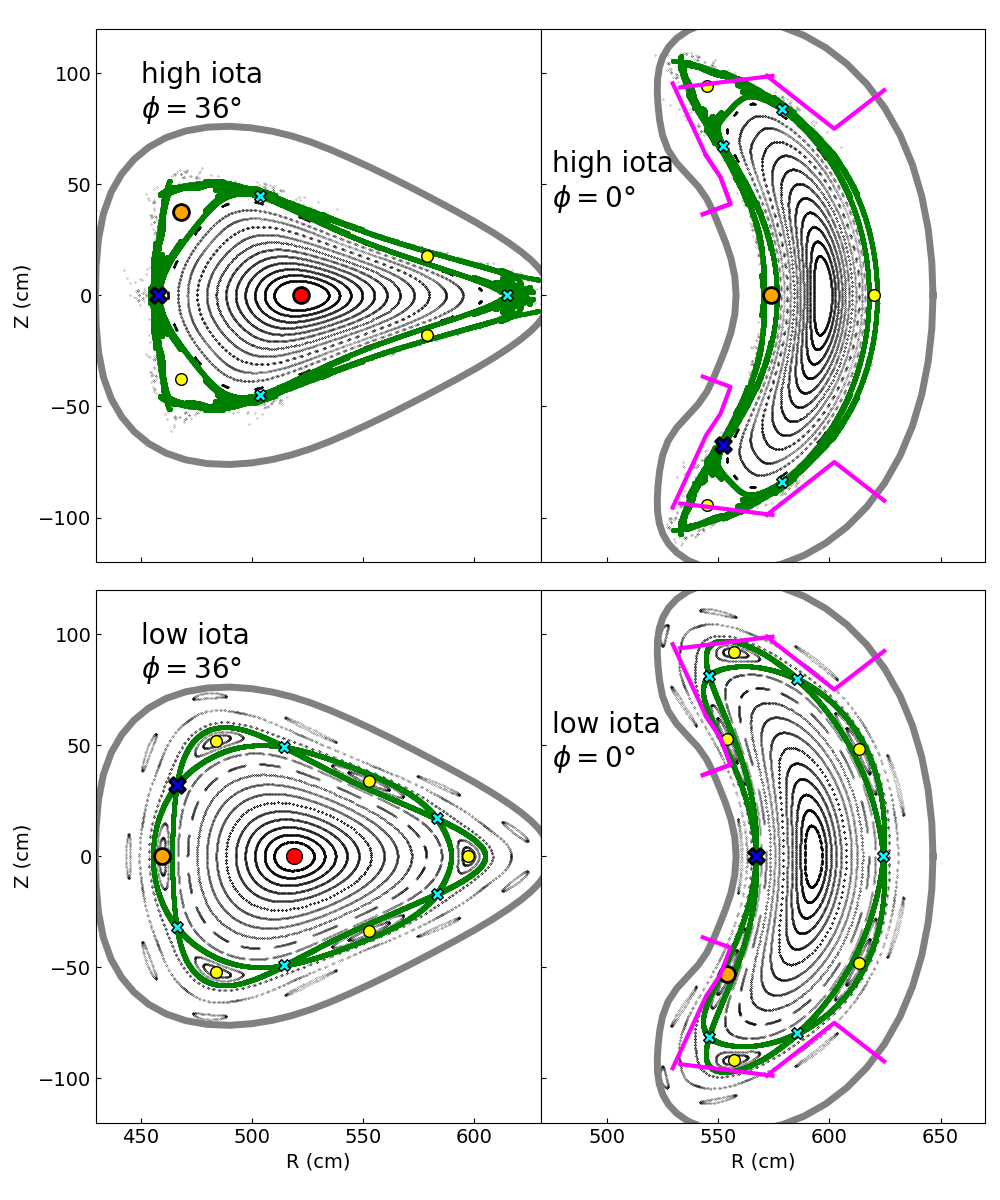}    
    \caption{A Poincar\'e section for the ``high iota" (upper row) and ``low iota" (lower row) configurations at $\phi=36^\circ$ (left) and $\phi=0^\circ$ (right).}
    \label{fig:low_high_iota_explanatory}
\end{figure}
Commercially viable stellarator reactors must have an exhaust solution, that is, the steady flow of heat and particles expelled from the confined region must be adequately managed e.g. 
Helium ash must be removed and plasma-facing components (PFCs) should not be thermally overloaded. The magnetic field in the stellarator edge plays an essential role in determining exhaust performance. Wendelstein 7-X (W7-X) is designed for an island divertor \cite{renner2004}, currently the most well-understood stellarator divertor concept and often used in stellarator reactor designs \cite{lion2025stellaris, bader2025power, goodman2025}. The island divertor relies on a magnetic island chain with a low-order rational rotational transform $\iota$ to divert the plasma. 
How the electromagnetic coils of W7-X influence its island chain(s) is the focus of this work.

W7-X has four varieties of electromagnetic coil \cite{andreeva2022magnetic, bozhenkov2016methods}, the first three of which are shown in figure \ref{fig:coils_plot}: (1) 5 unique ``non-planar" coil geometries, numbered 1-5 (referred to here as ``NPC 1", etc.); (2) 2 unique ``planar" coils, labelled ``A" and ``B" (``PC A", ``PC B"); (3) 1 unique ``control" (``CC", also called ``sweep") coil, which is close to the plasma (within the vacuum vessel) and carries a relatively small current; (4) ``trim" coils, which are furthest from the plasma and intended to only correct error fields. We do not study trim coils in this work. These coils can be independently powered, enabling a large range of unique magnetic configurations and hence a variety of magnetic topologies in the edge. Three paradigmatic configurations are the ``standard", ``high iota" and ``low iota" configurations. The standard configuration has an $\iota=m/n=5/5$ island chain in the edge (where $m$ and $n$ are the toroidal and poloidal periodicities of the island chain) which intersects the divertor plates, as shown in figure \ref{fig:standard_explanatory}. The high iota configuration has an edge $\iota=m/n=5/4$ island chain and low iota has an edge $\iota=m/n=5/6$ island chain, shown in figure \ref{fig:low_high_iota_explanatory}. A standard approach to model these island chains as a resonant magnetic perturbation of a hypothetical set of unperturbed nested flux surfaces\cite{bozhenkov2013service, bozhenkov2016methods}. Simple island models exist \cite{feng2022review} describing the relationship between island parameters (e.g. island width) and properties of the unperturbed surfaces (such as the magnetic shear) and the resonant perturbations. 

The role of coils on W7-X magnetic configurations and edge structures has been widely reported (e.g. \cite{andreeva2022magnetic}) and can be summarised as follows: (1) non-planar coils are responsible for the plasma shaping and generating the rotational transform $\iota$, and also control the mirror ratio; (2) the planar coils can add/remove toroidal magnetic field and vertical magnetic field, with the former shifting the $\iota$ profile up/down and the latter shifting the flux surfaces inwards/outwards (i.e. to larger/smaller major radius); (3) the control coil can control the island and phase of edge magnetic islands. This work seeks to quantify these effects over the operational space of W7-X, and to describe coil-island relationships beyond these principal roles.

Our focus is on fixed points of the field line map, most often X-points and O-points. Island chains consist of X-points and O-points and are therefore well-characterised by this technique. The main difference of our approach and the ``resonant perturbation model" is that we do not require the calculation of unperturbed nested surfaces (which in general is not trivial and solutions are non-unique). Here we only consider vacuum fields (the contribution of the plasma current is neglected), although plasma-generated fields have been shown to affect edge structures \cite{gao2019effects, killer2019effect, zhou2022equilibrium, knieps2021plasma}.

This work is organised as followed. We first (section \ref{sec:theory}) introduce theory relating to fixed points, in particular, we introduce the Jacobian $\MM$ of the field line map at a fixed point. In section \ref{sec:description_automated_scheme} we describe our fast automated scheme. We then perform two sets of coil current scans. The first, described in section \ref{sec:1d_scan}, is a 1D scan in each of the coils, taking the standard, high and low iota configurations as starting points. The second is a large scan of different combinations of coil currents. Section \ref{sec:summary_of_space} presents an overview of the the large scan and we explore three phenomena: the role of the planar coils in determining location and periodicity of fixed points (section \ref{sec:planar_coils}), the role of the control coil in modifying fixed point properties (section \ref{sec:control_coils}) and the factors governing the size of internal island chains (section \ref{sec:internal_islands}). Conclusions and outlook is presented in section \ref{sec:conclusions}.

\section{Relevant topological theory}\label{sec:theory}
Analysis of the field in a stellarator usually starts with constructing a Poncar\'e section \cite{poincare1893methodes}. 
This is usually done by choosing a plane of constant toroidal angle $\phi$ (using a cylindrical $(R, \phi, Z)$ coordinate system), and following magnetic field lines starting from a point $(R,Z)$ on the plane as they go around the device, noting each time the location $(R',Z')$ where that field line returns to the same plane. 
This also defines the Poincar\'e return map $f^n: (R, Z)\rightarrow (R', Z')$ which is the map defined by going $n$ times toroidally around the device. 
A Poincar\'e section is constructed by calculating the trajectories of dozens of field lines, for hundreds or thousands of toroidal transits around the device, and often illustrates well the structure of the magnetic field. 

There are two properties that stellarators (including W7-X) use to reduce the number of unique coil geometries they require, and thus simplify their construction. The first is \emph{field periodicity}: stellarators are constructed such that the field repeats itself $n_\text{fp}$ times in $\phi$ with a period $2\pi/n_\text{fp}(=72^\circ$ for W7-X since $n_\text{fp}=5$). 
Fewer unique coil geometries are required, they can just be copied and placed in their respective field periods. 
The second reduction is \emph{stellarator symmetry} \cite{dewar1998stellarator}. This implies that the magnetic field adheres to the relation $(B_R, B_\phi , B_Z) (R,\phi, Z) =  (-B_R, B_\phi , B_Z) (R, -\phi, -Z)$. 
This puts a strong relation between the coils at $\phi>0$ and $\phi<0$, which comes down to copying the coil at angle $+\phi$ to $-\phi$, rotating it upside-down, and changing the direction of the current. 
This further halves the number of unique coils. 
Because of field periodicity, the coils at $\phi<0$ re-appear in the second half of the field period, and stellarator symmetry implies the existence of a second symmetry plane halfway through the field period $\phi=\pi/n_\text{fp}$ ($=36^\circ$ for W7-X).

These symmetries also allow for faster computation of Poincar\'e sections. 
Field periodicity implies that every plane separated by $2\pi/n_{fp}$ is identical, and that we can iterate over $f^{1/n_\text{fp}}$. In addition, stellarator symmetry implies that the trajectory of a field line $(R(\phi), Z(\phi))$ started at $(R_0, 0, Z_0)$ and followed in the positive $\phi$ direction is related to the trajectory of a field line $(R'(\phi), Z'(\phi))$ started at $(R_0, 0, -Z_0)$ as follows: $(R'(-\phi), Z'(-\phi)) = (R(\phi), -Z(\phi))$. A consequence is that sections in the first half field period are related to the sections in the second half field period by a flip in the $Z=0$ plane, and are symmetric about $Z=0$ at the symmetry planes. Thus, all the unique information of a configuration can be presented by Poincar\'e sections between the first two symmetry planes (one half field period e.g. between $\phi=0$ and $\phi=36^\circ$ for W7-X). Here we assume that the magnetic field adheres strictly to field periodicity and stellarator symmetry, although small symmetry-breaking error fields can arise in the experiment and experimental configurations can deliberately break these symmetries \cite{bozhenkov2018measurements}. 

A Poincar\'e section contains special points, which are \emph{fixed} under the the Poincar\'e map $f$, such that $f^{n_\fixed/n_\text{fp}}(R,Z) = (R,Z)$.
Such a field line returns to the same location after $n_\fixed$ field periods ($n_\fixed/n_\text{fp}$ toroidal turns), and is thus a closed field line.  The magnetic axis is an example. It returns to the same $(R,Z)$ every period, and closes on itself after a full toroidal turn. 
Island chains contain fixed points, which return after multiple field periods.
The $5/5$ island chain of the standard configuration of W7-X contains (i.e. $n_\fixed=5$ fixed points), and the $5/4$ and a $5/6$ chains contain $\nfixed=4$ and $\nfixed=6$ fixed points respectively.

The structure of the map around a fixed point at $\xx_0$ can be understood from linearizing the field line map around it:
\begin{equation}\label{eq:linmap}
  f^n(\xx_0+\delta \xx)= f^n(\xx_0)+ \MM\cdot \delta\xx
\end{equation}
where $\MM$ (the Jacobian) is the linearized matrix of partial derivatives defined as: $\MM_{ij}=\partial_j f_i$, and $\delta \xx = \xx_i-\xx_0$ is a small step. 
This can also be written as:
\begin{align}
    R_f &= R_0 + M_{11}(R_i-R_0) + M_{12}(Z_i-Z_0) \label{eq:jacobian_1} \\ 
    Z_f &= Z_0 + M_{21}(R_i-R_0) + M_{22}(Z_i-Z_0) \label{eq:jacobian_2},
\end{align}
where $\xx_f = (R_f, Z_f)$ is the location of a field line started at $\xx_i=(R_i, Z_i)$ and traced $n_\fixed$ field periods and $\xx_0 = (R_0, Z_0)$. $\MM$ encapsulates several properties of the fixed point. Firstly, $\trm = M_{11} + M_{22}$ determines the ``flavour" of the fixed point (e.g. X-point, O-point), according to the following rules:\begin{enumerate}
    \item If $\trm >2$ the fixed point is an X-point.
    \item If $2>\trm >-2$ the fixed point is an O-point.
    \item $\trm =2$ fixed points include (but are not limited to) points on intact rational surfaces.
    \item If $\trm <-2$ the fixed point is a ``reflection hyperbolic fixed point". 
\end{enumerate}
The last ``flavour", probably least familiar to the reader, can occur in W7-X; however, they appear near the coils, and are typically inaccessible to the plasma (the plasma intersects the PFCs before interacting with the reflection hyperbolic (RH) point.) We therefore do not dwell on this phenomenon but refer the reader to \cite{lichtenberg1992chaotic, meiss1992symplectic, smiet2020bifurcations, smiet2025surfaces, davies2025topology}. 

For O-points, $\trm$ is also related to the rotational transform \textit{around the fixed point} with respect to the map according to $2\cos(2\pi \iota) = \trm$. For X-points, nearby field lines asymptotically approach or are exponentially repelled as the field lines are followed in the positive toroidal direction. The directions of attraction/repulsion are given by the ``stable"/``unstable" eigenvectors $\vv_\text{stable}, \vv_\text{unstable}$ of $\MM$ and the eigenvalues $\lambda_\text{stable}, \lambda_\text{unstable}$ describe the exponent of attraction/repulsion (related also to the Lyapunov exponent) \cite{wolf1986quantifying}. If one instead follows field lines in the negative toroidal direction, the directions of attraction/repulsion are swapped (unstable/stable eigenvectors are the directions of attraction/repulsion respectively.) The eigenvectors of the fixed point are dependent on the position along the trajectory of the fixed point (i.e. they vary with $\phi$) but the eigenvalues do not.

The last concept we introduce is the stable/unstable manifolds of the X-point: the set of (infinitely many) magnetic field lines which asymptotically approach the X-point when followed in the positive/negative toroidal direction. Near to the X-point, these lie along the stable/unstable eigenvectors. The manifolds tend to act as transport pathways for the plasma; for island divertors with negligible chaos (such as the standard configuration), the manifolds form the island ``separatrix", which diverts heat and particles away from the confined region. Of course, since the plasma can be transported in either direction along field lines, heat and particles are agnostic to whether a manifold is stable or unstable; ``stable" and ``unstable" should be understood as labels relating to a toroidal direction in which magnetic field lines approach a given X-point rather than as physically distinct concepts.

\section{Description of the scheme}\label{sec:description_automated_scheme}
We begin by finding the magnetic axis. We use a root finder to locate $(dR,dZ)=(R_f-R_0, Z_f-Z_0)=(0,0)$ points for the one field period map. For the initial guess we sample points along $Z=0$ at $\phi=36^\circ$, trace for one field period and select the ($R,Z$) point with smallest ($dR^2+dZ^2$). We calculate $\MM$ by tracing a set of nearby points and fitting $\{R_i, Z_i\}$ and $\{R_f, Z_f\}$ to equations \eqref{eq:jacobian_1} and \eqref{eq:jacobian_2} (see \ref{app:manifolds} for more detail).

We then seek fixed points with periodicities $n_\fixed=4,5,6$ i.e. $(dR,dZ)=(0,0)$ under the $n_\fixed$ field period map. We make multiple attempts using $N_\text{scan}=10$ points spaced along $Z=0$ in $\phi=36^\circ$ as starting points. Having found up to $N_\text{scan}$ fixed points, we remove duplicates, resulting in between $0$ and $N_\text{scan}$ unique fixed points in addition to the axis. 

If we have only found 1 unique fixed point in addition to the magnetic axis, it is likely that an island chain exists but the fixed point on the $\phi=36^\circ$ outboard midplane at is the same as that on the inboard midplane (for example the high and low iota configurations, see figure \ref{fig:low_high_iota_explanatory}). In this case we repeat the fixed point-finding scheme at $\phi=0^\circ$. Our approach is not guaranteed to find all fixed points, but it can be shown that for island chains, there must exist fixed points at $Z=0$ on both sides of the axis in both symmetry planes, and so at least some island fixed points are recovered. $\MM$ is calculated for each fixed point as for the magnetic axis.

\subsection{Implementation details using EMC3-Lite}
The field line tracing is performed by EMC3-Lite \cite{feng2022review}, which is interfaced via a Python ``wrapper". EMC3-Lite is a code designed for calculating PFC heat loads, but also includes a field line tracer which uses the Adams-Bashforth-Moulton fifth-order predictor-corrector scheme with fourth-order Runge-Kutta as the initial step. The magnetic field is calculated on a regular $(R,\phi,Z)$ grid using EMC3-Lite`s Biot-Savart solver\footnote{The magnetic field $(R,\phi,Z)$ grid is not to be confused with EMC3-Lite`s field-aligned grid \cite{feng2005simple}, which is used e.g. in the heat deposition calculation.}.

Two implementation details significantly speed up the calculations. Firstly, the magnetic field generated by each independent coil is calculated only once. The magnetic field for a given set of currents is then calculated as a linear sum of coil contributions. Secondly, the field line tracer is interfaced with the Python wrapper using f2py, allowing the magnetic field grids to be stored in memory, rather than reading the magnetic field from a file each time the field line tracer is called.

\section{Results of 1D scans}\label{sec:1d_scan}
\begin{figure}
    \centering
    \includegraphics[width=0.99\linewidth]{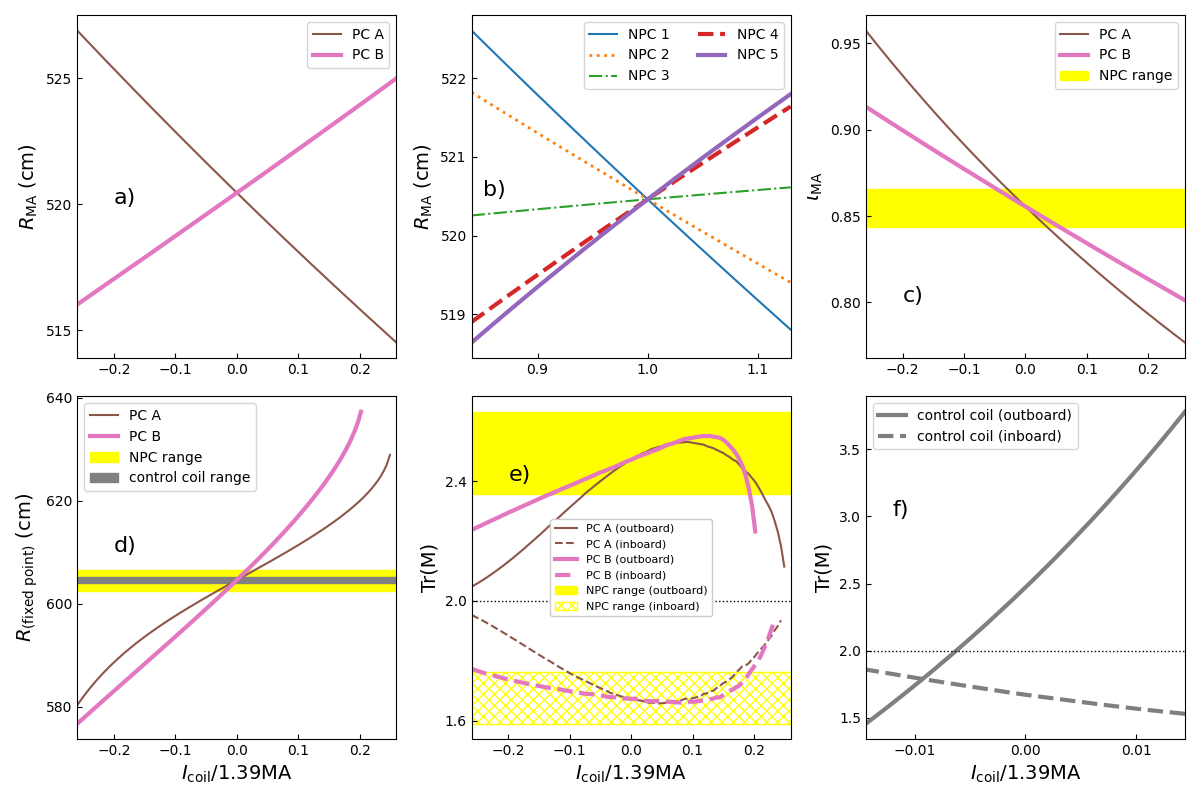}
    \caption{Coil currents scans with the standard configuration as the starting point. a) Magnetic axis location at $\phi=36^\circ$ ($R_\text{MA}$) as planar coils (PC) are varied. b) $R_\text{MA}$ variation with non-planar coils (NPCs). c) On-axis rotational transform $\iota_\text{MA}$ against PC currents. The range of $\iota_\text{MA}$ when NPC currents are scanned is shown as shaded yellow region. d) Location of $\phi=36^\circ$ outboard midplane fixed point $R_\fixed$ against PC currents, with NPC range shown as shaded yellow region and control coil (CC) range shown as shaded grey region. e) Jacobian trace $\trm$ for the $\phi=36^\circ$ outboard midplane fixed point (solid lines) and the $\phi=36^\circ$ inboard midplane fixed point (dashed line) against PC currents with NPC range shown in shaded yellow. f) $\trm$ against CC current for both fixed points.}
    \label{fig:1d_scans}
\end{figure}
\begin{figure}
    \centering
    \includegraphics[width=0.9\linewidth]{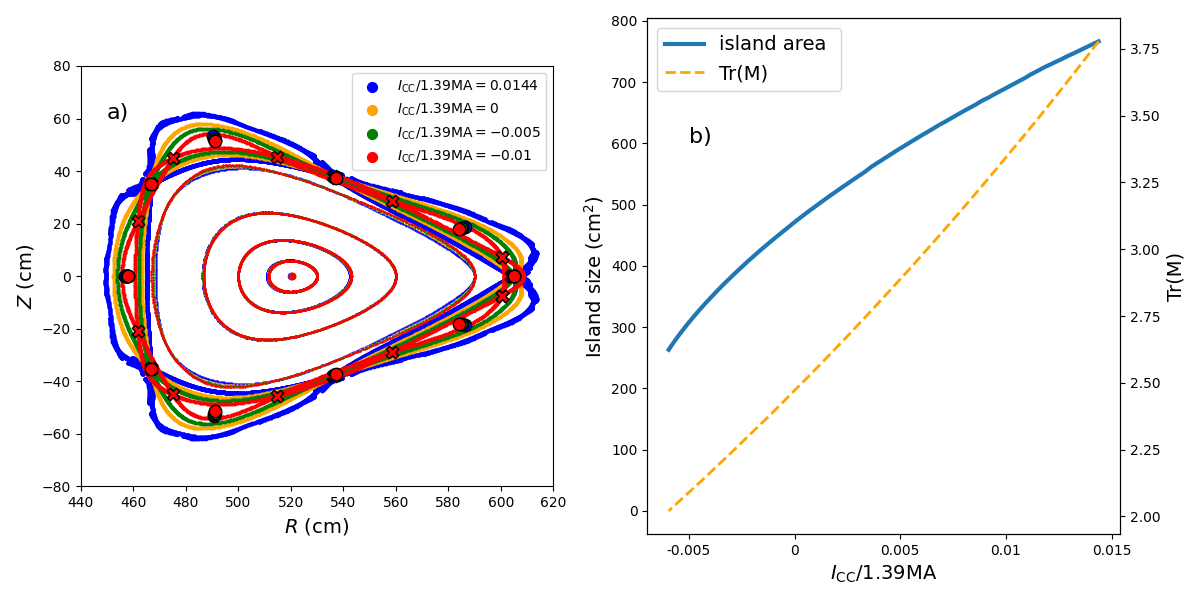}
    \caption{a) Poincar\'e sections, showing edge island chain as control coil current $I_\text{CC}$ is scanned. b) Island area and Jacobian trace $\trm$ as a function of $I_\text{CC}$.}
    \label{fig:cc_1d_scan}
\end{figure}
\begin{figure}
    \centering
    \includegraphics[width=0.98\linewidth]{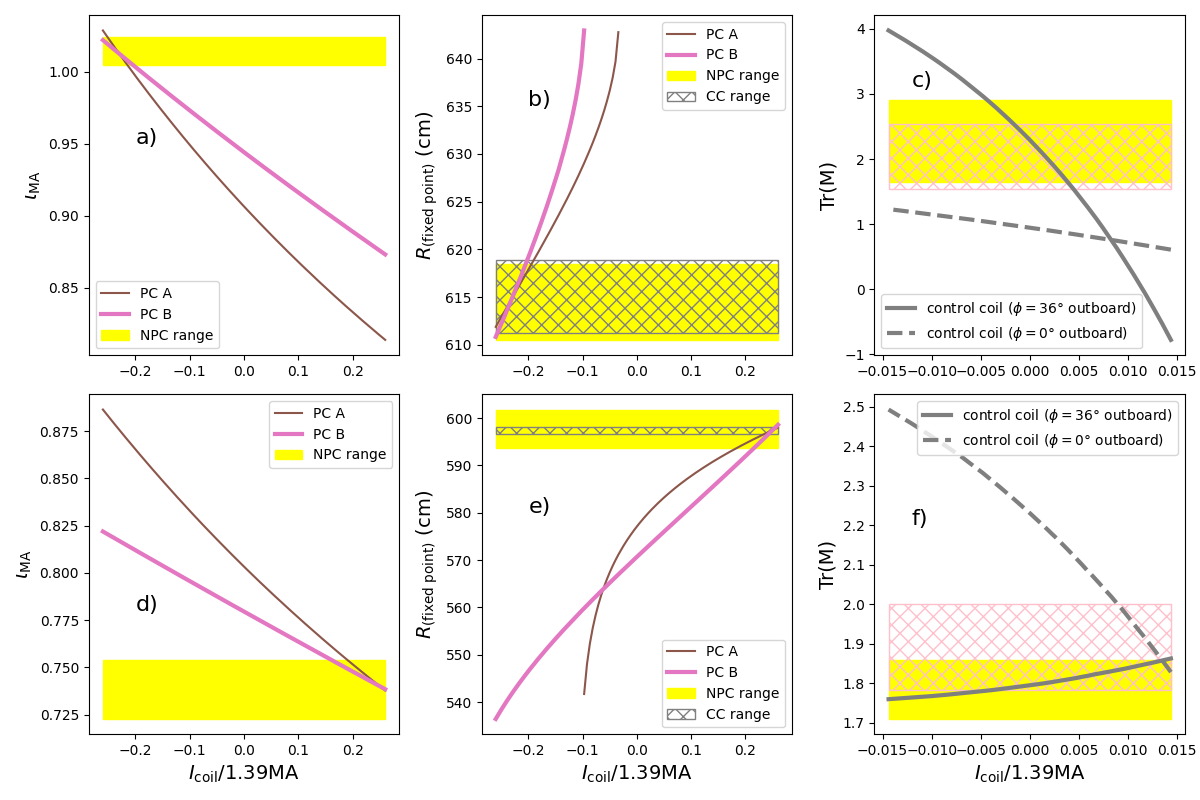}
    \caption{Coil currents scans with respect to the ``high iota" (plots a-c) and ``low iota" (d-f)  configurations. a) and d): $\iota_\text{MA}$ against PC current. b) and e): $R_\fixed$ against PC current with NPC and CC range shown as shaded areas. c) and f): $\trm$ for $\phi=36^\circ$ outboard midplane fixed point (solid line) and $\phi=0^\circ$ outboard midplane fixed point against CC current. }
    \label{fig:1d_scans_low_high_iota}
\end{figure}
We first perform a series of 1D scans using standard, high iota and low iota as starting points, by varying the current in each coil separately. The coil current $I_\text{X}$ reported here is the filament (i.e. total) current in coil X, equal to the winding current multiplied by the number of windings for the coil (which is $108$, $36$ and $8$ for the non-planar, planar and control respectively). The scan with respect to standard is shown in figure \ref{fig:1d_scans}. The $R$ location of magnetic axis at $\phi=36^\circ$ (labelled $R_\text{MA}$) can be moved by around $10$cm per planar coil and correlates with $(I_\text{PCA} - I_\text{PCB})$, and can also be moved by the non-planar coils by up to $4$cm (subplots a) and b)). $\iota_\text{MA}$ correlates with $(I_\text{PCA} + I_\text{PCB})$, with the non-planar coil currents having a smaller impact (subplot c)). The control coil has virtually no effect on either $R_\text{MA}$ (changing $R_\text{MA}$ by a maximum of $0.4$cm) or $\iota_\text{MA}$ (causing a maximum change of $0.003$). 

The variation in $R_\fixed$, the location of the fixed point on outboard midplane at $\phi=36^\circ$, is greatest for the planar coils (up to $60$cm), followed by the non-planar coils (up to $4$cm), and very little for the control coil (up to $1$cm) (subplot d)). However, the control coil has the strongest influence on $\trm$ for the X-point (a variation of more than $3$ over the experimentally relevant range), sufficient to achieve a X-point$\rightarrow$ O-point transition (indicated by $\trm$ crossing below 2 in subplot f)). Somewhat surprisingly, the O-point is far less affected by the $I_\text{CC}$.

Figure \ref{fig:cc_1d_scan} a) shows the island chain as $I_\text{CC}$ is scanned. The island separatrix (i.e. X-point manifolds) are calculated by sampling and tracing points along the eigenvectors of $\MM$ (see \ref{app:manifolds} for more information). Consistent with plot \ref{fig:1d_scans} f), the $5/5$ chain transitions to a $10/10$ chain, with an X$\rightarrow$O transition. Figure \ref{fig:cc_1d_scan} b) shows the variation in island area (here referring to the area of the island with an X-point at the outboard midplane at $\phi=36^\circ$, calculation described in \ref{app:island_area}) as the transition is approached. As $\trm \rightarrow 2$ for the X-point, the island area decreases but does not approach $0$. 

We repeat the coil current scans, starting from the high iota and low iota configurations, with selected results shown in figure \ref{fig:1d_scans_low_high_iota}. As in the ``standard" scan, the planar coils mostly determine the axis properties ($R_\text{MA}$ and $\iota_\text{MA}$) ($\iota_\text{MA}$ variation shown in subplots a) and d)) and the location of the fixed points $R_\fixed$ (subplot b)). 

However, several differences between the standard, high iota and low iota scans can be seen. For high iota, $\trm$ (shown in subplot c)) is more affected by all coil types (non-planar, planar and control) than for standard; all coil types can create an X$\rightarrow$O transition for the fixed point on the $\phi=36^\circ$ outboard midplane (i.e. a $4/4 \rightarrow 8/8$ island chain transition). For low iota, the fixed points are less affected by all coil types than standard (subplot f)). Another difference is that $\trm$ for the $\phi=36^\circ$ outboard midplane fixed point increases with $I_\text{CC}$ for high iota, the opposite trend to standard and low iota. Like standard, in low and high iota scans one fixed point in the island chain is more strongly affected by $I_\text{CC}$ than the other. For high iota, it is the $\phi=36^\circ$ outboard midplane fixed point, but for low iota it is the $\phi=0^\circ$ outboard midplane fixed point. A final difference is that for high iota, $I_\text{CC}$ can cause much larger variations in $R_\fixed$ than for standard or low iota (up to $8$cm).

An obvious reason for differences between these scans is the distance of the edge fixed points to the coils. The change to the magnetic field resulting from changes in coil currents is larger when the coil-fixed point distance is smaller, which qualitatively explains why $\trm$ variation is greatest for high iota (because $R_\fixed$ is largest, i.e. the fixed points are closer to the coils) and smallest for low iota. Another difference is the $\iota$ of the fixed points, which determines the poloidal/toroidal space which the fixed point trajectories sample. This might explain the differences in $\trm (I_\text{CC})$ trends, although we do have a quantitative or predictive model for these variations.

\section{Summary of the large scan}\label{sec:summary_of_space}
\begin{figure}
    \centering
    \includegraphics[width=0.99\linewidth]{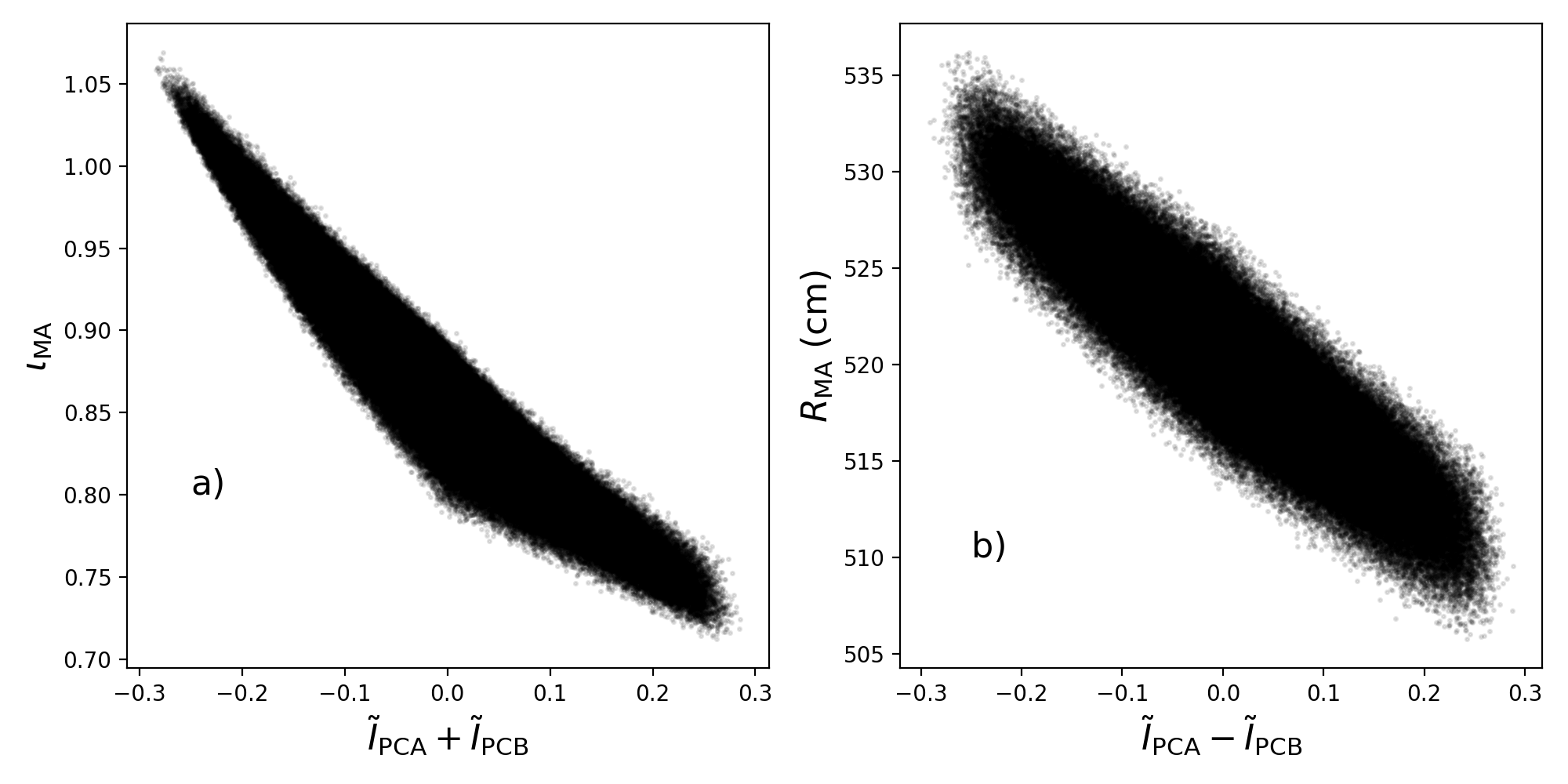}
    \caption{a) On-axis rotational transform against normalised planar coil currents $(\pcanorm+\pcbnorm)$. Currents normalised to mean non-planar coil current. b) Location of magnetic axis at $\phi=36^\circ$ against $(\pcanorm-\pcbnorm)$.}
    \label{fig:magnetic_axis_properties}
\end{figure}
\begin{figure}
    \centering
    \includegraphics[width=0.48\linewidth]{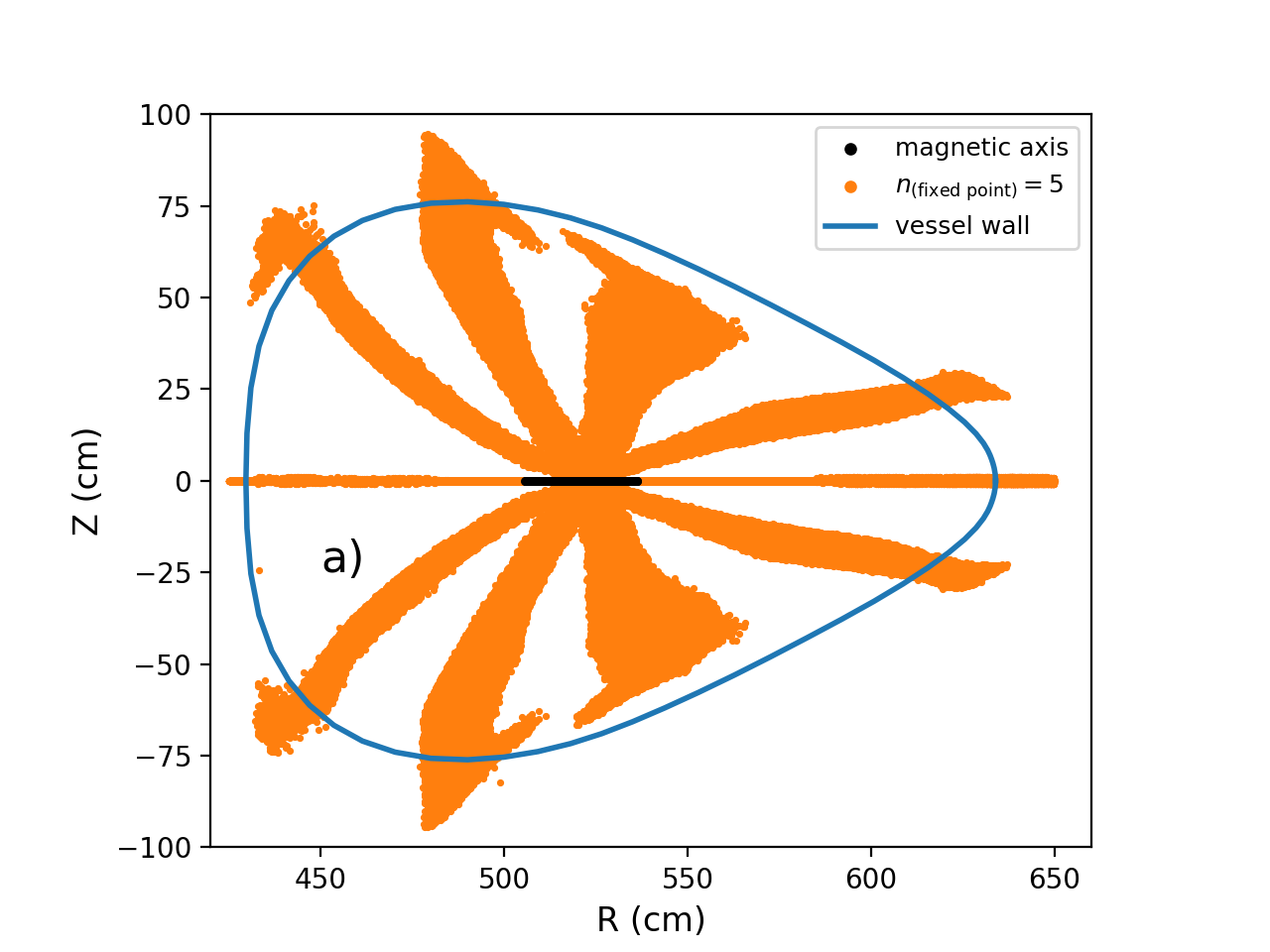}
    \includegraphics[width=0.48\linewidth]{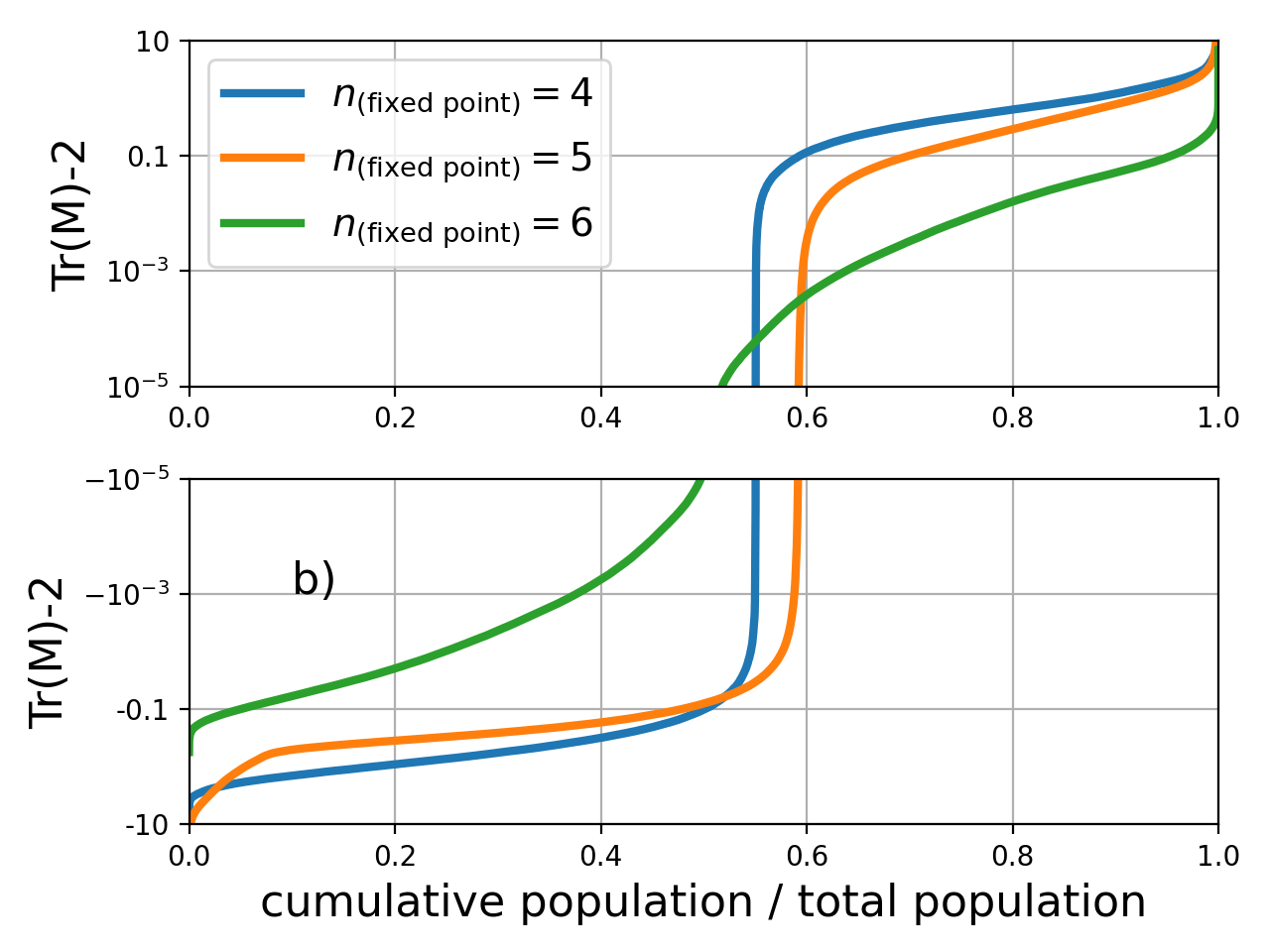}    
    \caption{a) Illustration of all $\nfixed=5$ found in our large coil current scan. b) $\trm$ of all fixed points found in the large scan, plotted in ascending order.}
    \label{fig:summary_edge_fps}
\end{figure}
We analyse over $2.8\times 10^5$ different coil current combinations, with currents for each coil selected uniformly randomly within the prescribed range (see \ref{app:constraints_hyperparameters} for details). We perform the scan using a single CPU with an average computational cost of $0.26$ CPU-seconds per configuration.

The magnetic axis location at $\phi=36^\circ$ as a function of $(\pcanorm-\pcbnorm)$ (the parameter controlling inward/outward shift) is shown in figure \ref{fig:magnetic_axis_properties}. The tilde ( $\tilde{}$ ) denotes normalised current, defined as $\tilde{I}_X = I_X \cdot 5/(I_\text{NPC1}+I_\text{NPC2}+I_\text{NPC3}+I_\text{NPC4}+I_\text{NPC5})$ for coil $X$, i.e. the coil current divided by the average NPC current. $R_\text{MA}$ spans the range $R=(505, 535)$ and has a strong (approximately linear) correlation with $(\pcanorm-\pcbnorm)$. $\iota_\text{MA}$ (figure \ref{fig:magnetic_axis_properties} b)) strongly correlates with  $(\pcanorm+\pcbnorm)$ and spans the range $0.71$ - $1.07$.

The $\nfixed=5$ fixed points are represented in figure \ref{fig:summary_edge_fps}. $10$ distinct ``bands" can be seen, with O-, X- and RH-points present in each band. Fixed points living outside these bands (for example the $20/20$ island chain) are possible, but would not be found by our algorithm. $\nfixed=5$ fixed points are present at all minor radii (which is not the case for the $\nfixed=4$ and $\nfixed=6$). 

$\trm$ for all edge fixed points found by our scheme is shown in figure \ref{fig:summary_edge_fps} b). $\trm$ varies the most for the $\nfixed=4$ and least for $\nfixed=6$; the middle $80\%$ of points (i.e. between 0.1 and 0.9 in plot \ref{fig:summary_edge_fps} b)) spans the range $0.58-3.20$, $1.50-2.78$, $1.94-2.05$ for the $\nfixed=4$, $5$ and $6$ respectively.

\subsection{Role of planar coils}\label{sec:planar_coils}
\begin{figure}
    \centering
    \includegraphics[width=0.98\linewidth]{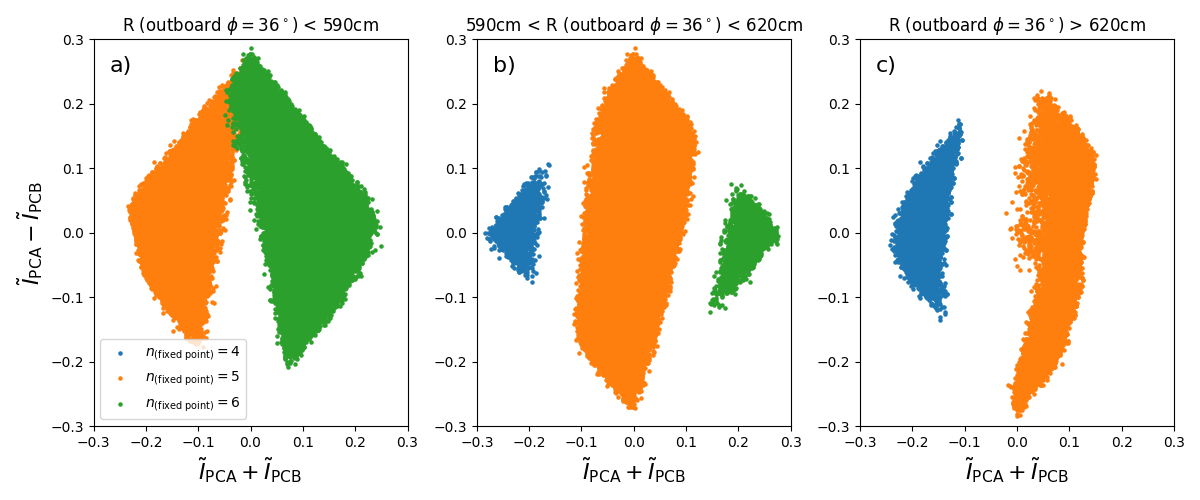}       
    \caption{Fixed point occurrence as a function of planar coil currents, categorised by location of $\phi=36^\circ$ at outboard midplane fixed point. a) ``internal" fixed points. b) ``divertor-relevant" fixed points. c) ``Near-coil" fixed points.}
    \label{fig:fixed_points_vs_pc}
\end{figure}
We loosely define three categories of fixed points, according to their spatial position: (1) ``internal" fixed points, defined as those for which $R<590$cm for the $\phi=36^\circ$ outboard midplane fixed point. These are likely to form internal island chains i.e. the island chains do not intersect PFCs; (2) ``divertor-relevant" fixed points,  $590 < R <620$cm for the $\phi=36^\circ$ outboard midplane fixed point, which are likely to form PFC-intersecting island chains; (3) ``near-coil" fixed points, $R>620$cm, which are less likely to have experimental relevance since the structures they form are likely to be shielded from the plasma by PFCs. We emphasise that this is a loose categorisation.

The distribution of fixed points with respect to $(\pcanorm + \pcanorm)$ and $(\pcanorm - \pcanorm)$ in these three categories are shown in figure \ref{fig:fixed_points_vs_pc}. The overall shape of the phase space of $(\pcanorm + \pcanorm)$ and $(\pcanorm - \pcanorm)$ is a diamond. Within the diamond, fixed points of each periodicity have a well-defined position in phase space, mostly determined by $(\pcanorm + \pcanorm)$. For ``divertor-relevant" fixed points, all three periodicities are present, with $\nfixed=5$ points possible over the full range of $(\pcanorm - \pcanorm)$. Only $\nfixed=5$ and $6$ ``internal" fixed points are possible with our selection of current limits and only $\nfixed=4$ and $5$ ``near coil" fixed points are possible.

\subsection{Role of control coil}\label{sec:control_coils}
\begin{figure}
    \centering
    \includegraphics[width=0.9\linewidth]{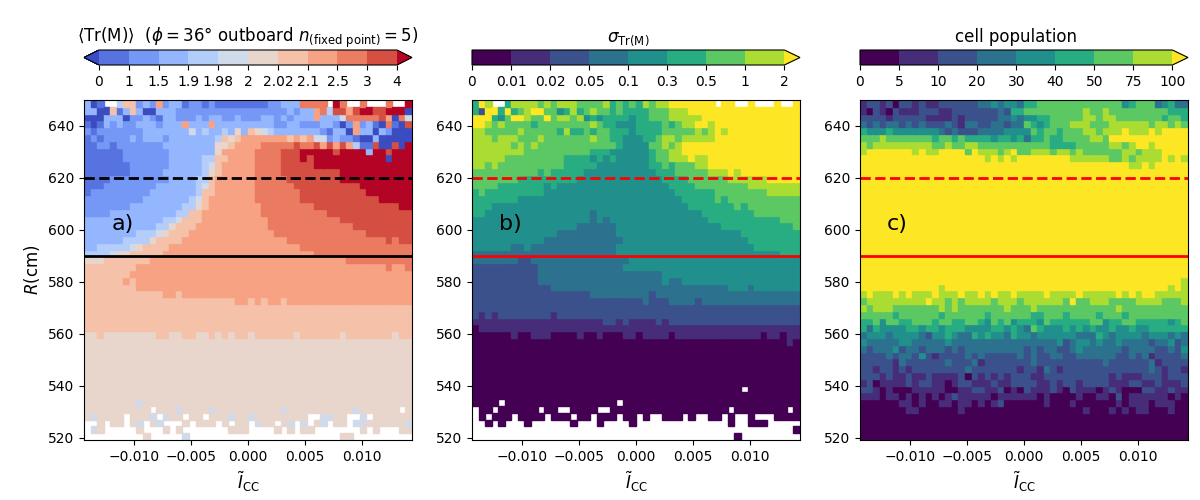}    
    \includegraphics[width=0.9\linewidth]{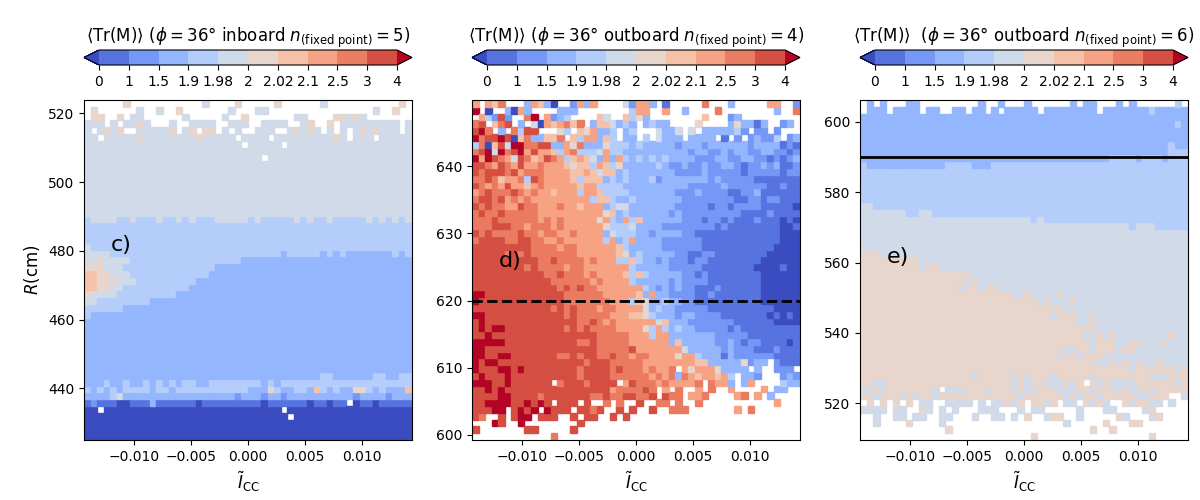}    
    \caption{Statistics of $\trm$ for fixed points against control coil current and $R$. Upper row:  $\nfixed=5$ $\phi=36^\circ$ outboard midplane fixed points, showing a) $\langle\trm\rangle$, b) standard deviation and c) cell population (right) for binned data. Lower row: $\langle\trm\rangle$ for d) $\phi=36^\circ$ inboard midplane $\nfixed=5$ fixed points, e)  $\phi=36^\circ$ outboard midplane $\nfixed=4$ and f) $\phi=36^\circ$ outboard midplane $\nfixed=6$ (right).}
    \label{fig:fixed_points_vs_R_and_cc}
\end{figure}
Section \ref{sec:1d_scan} shows that $\ccnorm$ has a greater effect on $\trm$ for fixed points located at larger minor radius, but also that $\ccnorm$ can unequally affect different fixed points in a single island chain. This is supported by our large scan. Figure \ref{fig:fixed_points_vs_R_and_cc} a)-d) characterises $\trm (R, \ccnorm) $ for $\nfixed=5$. The data is binned and we describe a given $(R, \ccnorm)$ cell by a mean $\trm$ (denoted $\langle\trm\rangle$), a standard deviation $\sigma_{\trm}$ and a cell population $N$. Figure \ref{fig:fixed_points_vs_R_and_cc} a) shows $\langle\trm\rangle$ for $\phi=36^\circ$ outboard midplane fixed points. Within the ``divertor relevant" range (between the dashed and solid horizontal lines), $\langle\trm\rangle$ increases with $\ccnorm$ and the magnitude of $(\langle\trm\rangle -2 )$ increases with $R_\fixed$. $\sigma_{\trm}$ increases with $R_\fixed$ (subplot b)), often exceeding 2 in the ``near coil" region. Subplot c) shows that $\nfixed=5$ mostly occupy the ``divertor-relevant" range.

Figure \ref{fig:fixed_points_vs_R_and_cc} d) shows $\langle\trm\rangle$ for the $\nfixed=5$ on the $\phi=36^\circ$ inboard midplane. As in section \ref{sec:1d_scan}, there is little dependence of $\trm$ on $\ccnorm$. $\vert (\langle\trm\rangle -2 ) \vert $ increases with distance from the magnetic axis. A sharp bifurcation occurs around $R=455$cm, with $\vert (\langle\trm\rangle -2 ) \vert $ becoming large and negative. This bifurcation is curious, but is likely to be in the ``near-coil" region and is not explored here.

Figure \ref{fig:fixed_points_vs_R_and_cc} e) and f) show $\langle\trm\rangle$ for $\phi=36^\circ$ outboard midplane fixed points with $\nfixed=4$ and $6$. For $\nfixed=4$, a strong correlation exists between $\langle\trm\rangle$ and $\ccnorm$, with the opposite trend to $\nfixed=5$. There is little dependence for $\nfixed=6$, and $\vert (\langle\trm\rangle -2 ) \vert $ tends to be small, but increases with distance from magnetic axis. This also supports the findings of section \ref{sec:1d_scan}.

Finally, we note that $\nfixed=5$ and $6$ have a clear preferred phase, with the majority (85\%) of $\nfixed=6$ $\phi=36^\circ$ outboard midplane fixed points being O-points, the majority (78\%) of such $\nfixed=5$ being X-points. In contrast, around half (53\%) of $\nfixed=4$ $\phi=36^\circ$ outboard midplane fixed points are X-points (no clearly preferred phase).

\subsection{Island size for ``internal" islands}\label{sec:internal_islands}
\begin{figure}
    \centering
    \includegraphics[width=0.99\linewidth]{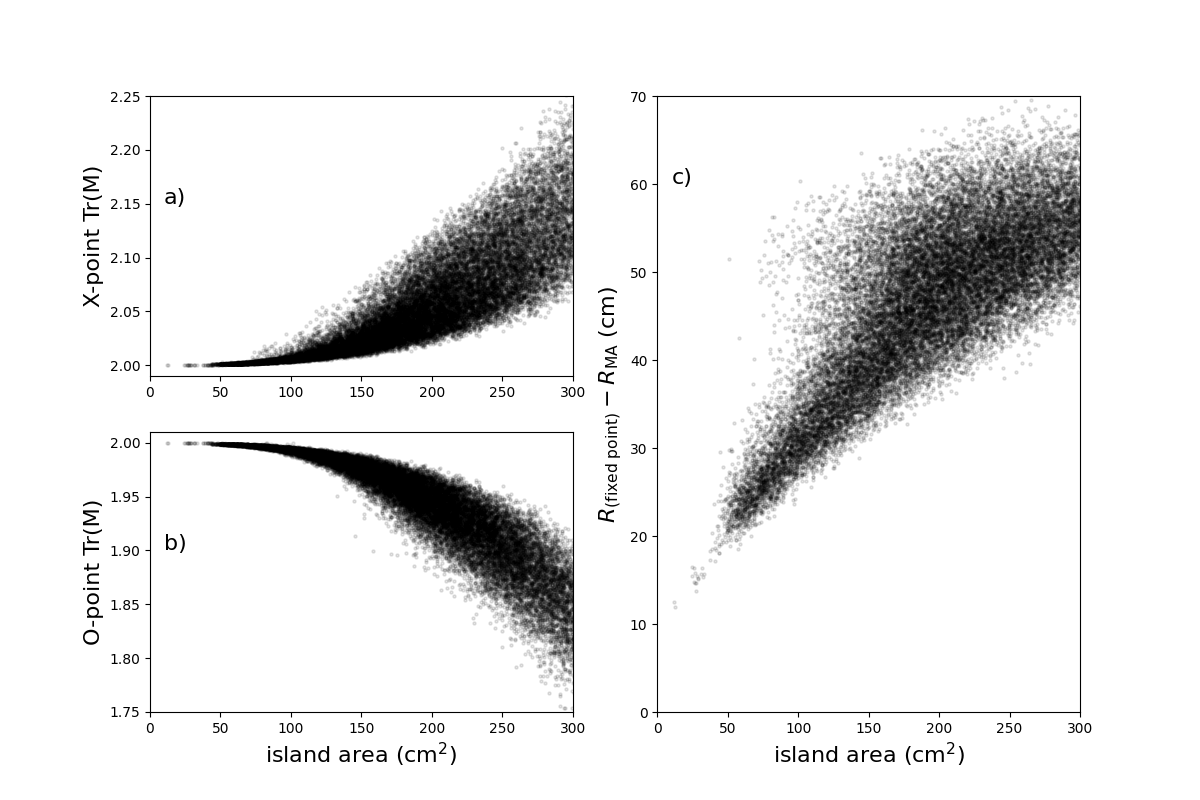}     
    \caption{a) and b) Comparison between fixed point $\trm$ and island area for ``internal" 5/5 island chains. c) Correlation between island area and minor radius of island chain.}
    \label{fig:proxies}
\end{figure}
Section \ref{sec:control_coils} shows that $\trm$ for fixed points tends to approach $2$ as the minor radius of the fixed point decreases, and section \ref{sec:1d_scan} shows that $\trm \rightarrow 2$ correlates with decreasing island size. 
This is indirectly what is targeted when the Cary and Hanson method~\cite{cary1986stochasticity} for stochasticity reduction in stellarators is used. 
Here a quantity called Greene's residue~\cite{greene1968two, mackay1992greene}, calculated through $\mathcal{R}=\tfrac{1}{2} - \tfrac{1}{4}\trm$ is minimized to reduce and eliminate internal islands. 
This quantity of course goes to zero when $\trm$ approaches 2. 
This is used to eliminate internal islands, which have historically been considered undesirable for fusion reactors (since they flatten the pressure profile across the island). However, recent experimental results \cite{andreeva2022magnetic} have found that island chains within the confined region but near the edge appear to suppress MHD activity and enable an increase in volume-normalised diamagnetic energy. The means to tailor internal island chains could therefore be useful for preparing future experiments.

To address this, we compare island size and $\trm$ for the $\nfixed=5$ ``internal" fixed points, shown in figure \ref{fig:proxies}. For reasons described in \ref{app:island_area}, the area calculation may fail, and our approach is limited to configurations with X-points on the $\phi=36^\circ$ outboard midplane and O-points on the $\phi=36^\circ$ inboard midplane. With these caveats, we calculate an island area for $32,949$ configurations (the total number of configurations with $\nfixed=5$ internal fixed points is $25,276$). We find that island area correlates positively with $\vert \trm-2 \vert$ for both the X- and O-points of the island chain, albeit with a large spread. 
Optimisations based on Greene`s Residue can therefore remove islands by minimizing this residue, but if a specific island size is desired, other metrics should be used (such as~\cite{geraldini2021adjoint}.) 
We also find that island area increases with the minor radius ($R_\fixed - R_\text{MA}$) (figure \ref{fig:proxies} c)). This might be expected, since the island area for a given island width should increases approximately linearly with minor radius. 

\section{Conclusions}\label{sec:conclusions}
This work presents a fast automated method to characterise fixed points in vacuum stellarator-symmetric magnetic fields, and applies this W7-X. We restrict our study to edge fixed points with periodicity of $\nfixed=4$, $5$ and $6$ field periods, but there is no reason the algorithm cannot be applied beyond this. 

We perform scans where each coil current is individually varied, relative to the ``standard", ``high iota", and ``low iota" configurations, and also a large scan in which all coil currents are varied simultaneously. These confirm the following behaviours: (1) the non-planar coils, responsible for generating rotational transform and plasma shaping, have relatively little effect on magnetic axis location and rotational transform or on edge fixed points; (2) the planar coils control inward and outward shift of the plasma and can boost the $\iota$ profile up and down, and thus principally control the periodicity and location of edge fixed points; (3) the control coil has a relatively large effect on fixed points in the edge, with the sensitivity of fixed points to coil current increasing with the minor radius of the fixed points. These are well-expected \cite{andreeva2022magnetic}, but for the first time are quantified over a large scan. 

We also find that the control coil has a surprisingly unequal effect on the X- and O-points within an island chain, which (for example) causes the $5/5$ island chain in the standard configuration to transition to a $10/10$ island chain: the X-point is changed to an O-point but the O-point experiences little change. Understanding the reasons for this would be a useful piece of future work, and possible methods to address these have been recently published \cite{hudson2025sensitivity, wei2025shifts}. Preliminary investigations \cite{daviesinpreparation} indicate that the changes to $\trm$ arise from highly toroidally and poloidally localised changes to the magnetic field induced by the control coil. The implications of this for transport in the edge (compared with simple island models in which the island rotational transform is uniform with toroidal angle) is unknown. Our final result is finding a strong correlation between $\vert \trm-2 \vert$ and island area for ``internal" island chains (i.e. not intersecting plasma-facing components), which may help to select experimental candidates with internal island chains of a desired width.

\section{Acknowledgements}
This work has been carried out within the framework of the EUROfusion Consortium, partially funded by the European Union via the Euratom Research and Training Programme (Grant Agreement No 101052200 — EUROfusion). The Swiss contribution to this work has been funded by the Swiss State Secretariat for Education, Research and Innovation (SERI). Views and opinions expressed are however those of the author(s) only and do not necessarily reflect those of the European Union, the European Commission or SERI. Neither the European Union nor the European Commission nor SERI can be held responsible for them. This work was supported by a grant from the Simons Foundation (1013657, JL).

\section*{References}
\bibliographystyle{vancouver}
\bibliography{references}

\begin{appendix}

\section{Illustration of calculation of Jacobian}\label{app:manifolds} 
\begin{figure}
    \centering
    \includegraphics[width=0.8\linewidth]{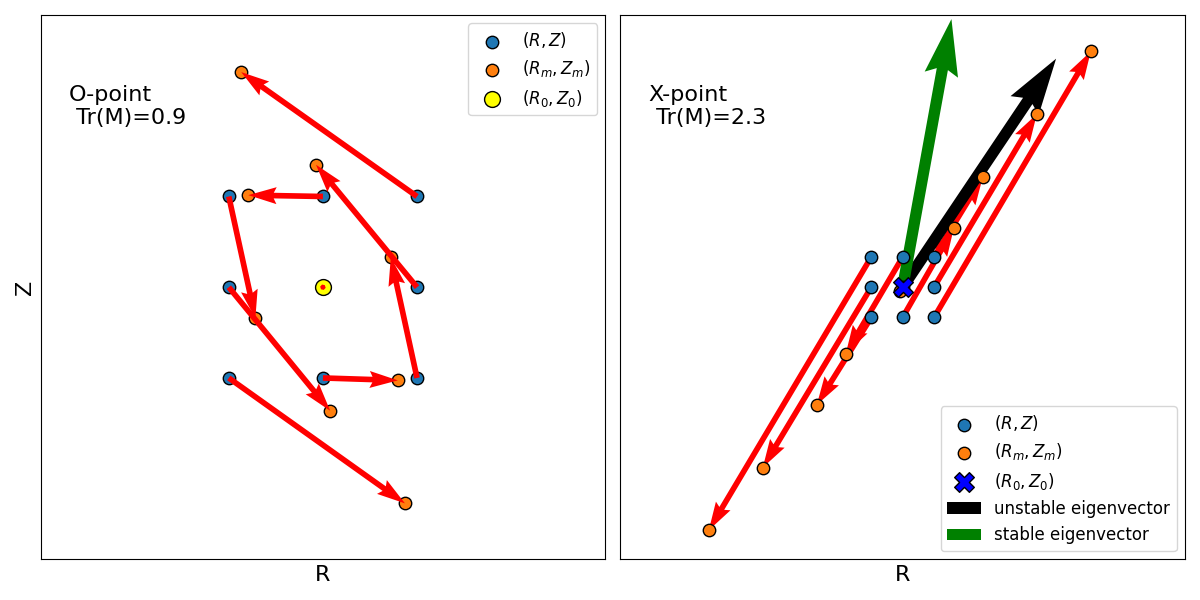}
    \caption{
    Illustration of calculating the Jacobian $\MM$ of the Poincar\'e map for fixed points, using an initial grid of points (blue circles) and their image under the field line map (orange circles). Left: O-point, for which $\trm<2$. Right: X-point, for which $\trm>2$. The eigenvectors of $\MM$ for the X-point are shown as black and green arrows.}
    \label{fig:explanatory_high_iota_a}
\end{figure}
The calculation of the Jacobian $\MM$ is shown in figure \ref{fig:explanatory_high_iota_a}. We construct a $3\times 3$ grid of points around the given fixed point and trace for $\nfixed$ field periods, and fit the points to equations \eqref{eq:jacobian_1} and \eqref{eq:jacobian_2}. The eigenvectors of $\MM$ for the X-point are easily calculated, and sampling points on the eigenvectors recovers the manifolds of the X-point (the island separatrix for a regular non-chaotic island chain).

\section{Island area calculation}\label{app:island_area}
\begin{figure}
    \centering
    \includegraphics[width=0.8\linewidth]{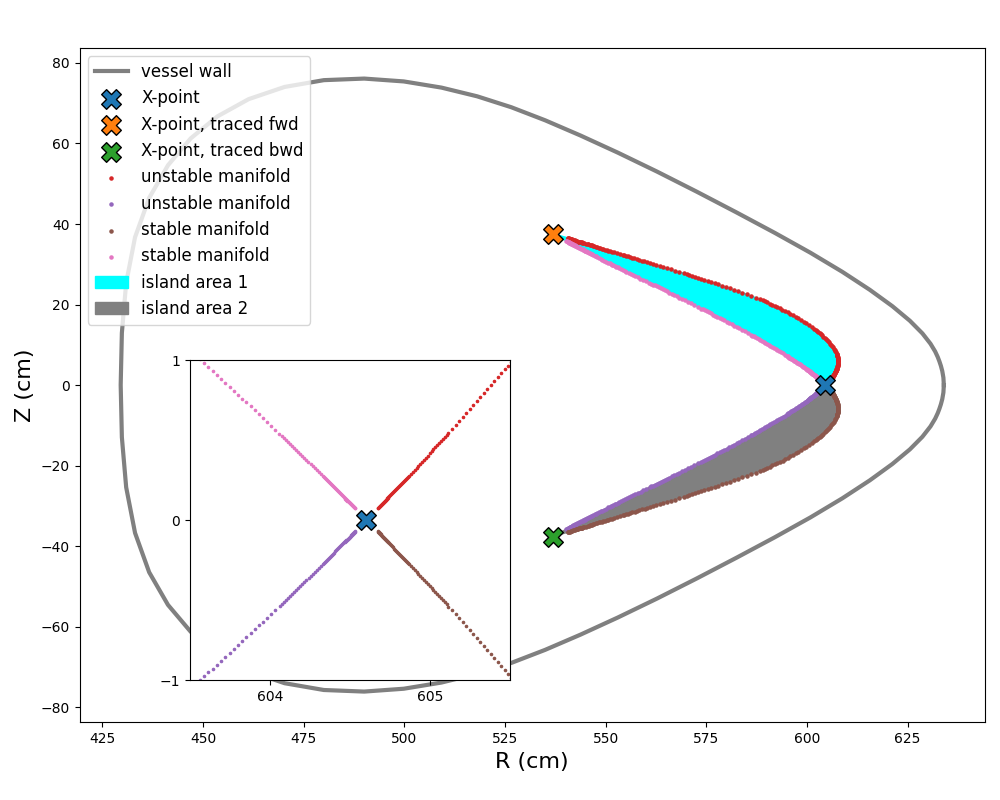}
    \caption{
    Illustration of the island area calculation scheme, applied to the standard configuration.}
    \label{fig:area_calculation_illustration}
\end{figure}
The island area calculations presented here are only performed for configurations with $\nfixed=5$ fixed points, and only for configurations with an X-point on the $\phi=36^\circ$ outboard midplane and an O-point on the $\phi=36^\circ$ inboard midplane, which includes (for example) the standard configuration (see figure \ref{fig:standard_explanatory}). We approximate the area by finding the eigenvectors of $\MM$ for the X-point and tracing the stable and unstable manifolds. If these manifolds become sufficiently close to the forward image of the X-point (within a distance $d_\text{tol}=0.02d_X$, where $d_X$ is the distance from the X-point to its forward image), we use the manifolds to construct a closed polygon, for which the area is calculated using Gauss's shoelace algorithm. As a check against numerical error, we repeat the calculation for the backward image of the X-point and compare the calculated areas. If the areas differ too greatly (by more than $A_\text{tol}=10\text{cm}^2$), or if the manifolds do not sufficiently approach the image of the X-point (which might happen for example if the island chain has $m/n=15/15$), the scheme is considered to have failed. An example of the calculation is shown for the standard configuration in figure \ref{fig:area_calculation_illustration}.

It is worth noting that the area of the island varies with toroidal position of the island. Calculating instead the toroidal magnetic flux in the island (for example) would create a toroidally invariant ``island size", although the toroidal flux should then be normalised to a reference magnetic field strength because field strength (and hence toroidal flux) can be arbitrarily scaled without affecting the island geometry.

\section{Selection of coil current constraints}\label{app:constraints_hyperparameters}
Experimentally, the coil currents in W7-X are constrained by a number of factors. These include: the electrical limit on the coil winding current, the constraints on the mechanical forces permitted by the machine, and the requirement that the magnetic field strength allows the plasma to couple to the ECRH beams. Because rigorously applying all of these constraints is relatively difficult, we use simple constraints the limits, with non-planar coil filament currents restricted to the range ($1.168$ - $1.57$MA), planar coil filament currents to the range ($-0.361$ - $+0.361$MA) and control coil filament current to ($-20$ - $+20$kA). These are within the normal range of coil currents operated by W7-X \cite{andreeva2022magnetic}.
\end{appendix}
\end{document}